\documentclass[preprint,12pt]{elsarticle}

\usepackage{graphicx,dcolumn,bm,xcolor,lineno,ulem,amsmath,amssymb}
\usepackage[utf8]{inputenc}
\usepackage{booktabs} 
\usepackage{comment}
\usepackage{makecell}
\usepackage{multirow}
\usepackage{tabularx}

\journal{Chaos, Solitons \& Fractals}

\begin{document}

\begin{frontmatter}

\title{Ranking football teams via the higher-order decomposition of performance networks} 

\author[aff1,aff2]{Andres Chacoma\corref{cor1}}
\ead{achacoma@df.uba.ar}

\author[aff3,aff4]{Juan I. Perotti}

\author[aff3,aff4]{Orlando V. Billoni}

\cortext[cor1]{Corresponding author}

\address[aff1]{Universidad de Buenos Aires, Facultad de Ciencias Exactas y Naturales, Departamento de F\'{i}sica, Buenos Aires, Argentina}

\address[aff2]{CONICET -- Universidad de Buenos Aires, Instituto de F\'{i}sica Interdisciplinaria y Aplicada (INFINA), Buenos Aires, Argentina}

\address[aff3]{Universidad Nacional de C\'{o}rdoba, Facultad de Matem\'{a}tica, Astronom\'{i}a, F\'{i}sica y Computaci\'{o}n, C\'{o}rdoba, Argentina}

\address[aff4]{CONICET -- Universidad Nacional de C\'{o}rdoba, Instituto de F\'{i}sica Enrique Gaviola (IFEG), C\'{o}rdoba, Argentina}

\begin{abstract}
We propose a unified methodological framework to quantify team performance in elite football by combining event-level performance metrics, higher-order network representations, and algebraic ranking methods. Using data from the 2017--2018 season of the five major European leagues, we construct metric-specific weighted graphs in which teams are connected through relative performance indicators. These graphs are analyzed via Hodge decomposition, and the gradient component is used to derive metric-based team ratings. The resulting rankings are systematically compared with the true league standings using Pearson and Kendall correlation measures, revealing strong metric- and league-dependent effects. 
Furthermore, by analyzing the ratio between solenoidal and total flow energies, we show that local cyclic dynamics structurally limit the gradient component's capacity to reconstruct the ranking. This topological inconsistency acts as a structural fingerprint of each league's ``competition style'' successfully mapping the studied systems into distinct regimes: highly hierarchical structures (England and Italy), tactical parity driven by generalized loops (Germany), and pockets of localized chaos (France and Spain). 
Lastly, we introduce a composite rating obtained as a parsimonious linear combination of metric-based ratings, optimized separately for each league. This composite approach significantly improves predictive power and allows the relative importance of different performance indicators to be quantified in a league-specific manner. Our results demonstrate how higher-order network methods provide a flexible and interpretable framework to uncover latent performance structures in football, offering a complementary perspective to outcome-based rankings and a general approach applicable to other oppositional sports.
\end{abstract}


\end{frontmatter}

\section{Introduction}

In recent years, the study of sports competitions has undergone a significant transformation through the adoption of theoretical frameworks drawn from complexity science
\cite{barthelemy2025fragility,Clauset2015safe,buldu2019defining,Ribeiro2012anomalous,Chacoma2022simple,Chacoma2023probabilistic,chacoma2025emergent,zappala2022role,ibanez2018relative,galeano2022using}. 
Sports championships, characterized by the dynamic interaction of multiple agents (players, teams, contexts), nonlinear outcomes, and the emergence of non-trivial collective patterns, are increasingly understood as complex adaptive systems \cite{silva2016sports}. This perspective makes it possible to move beyond purely descriptive statistical analyses and to address fundamental questions regarding competitive dynamics, performance evolution, and the underlying structure of play from robust theoretical principles.

This intersection between complex systems and sport is fertile from a dual perspective. From an academic standpoint, it offers a unique laboratory for validating and developing theories of networks, competitive dynamics, diffusion processes, and the emergence of hierarchies in social systems governed by clearly defined rules. From the high-performance perspective, the quantitative and objective modeling of competition is currently transforming decision-making processes \cite{ashford2021understanding}. Providing coaches, analysts, and decision-makers with data-driven tools to assess performance beyond immediate outcomes, diagnose structural strengths and weaknesses, and optimize strategies has become a priority in the pursuit of sustainable competitive advantages.

While this approach has been successfully applied to various team and individual sports, football stands out as a particularly rich domain. 
Its global popularity, the growing availability of high-resolution data (event data, tracking data), and the inherent tactical complexity of a continuous-flow, low-scoring game make it an ideal case study. Recent research has employed concepts from complex systems to model ball possession as a diffusion process, characterize collective creativity, and identify critical phases of play, demonstrating the considerable potential of this theoretical framework \cite{martinez2020spatial,cao2023football,yamamoto2024theory,chacoma2020modeling,chacoma2021stochastic,chacoma2025data,amichay2025characterizing}.

Within the methodological toolkit of complex systems, complex network theory has emerged as a particularly powerful approach for football analysis. By modeling players or teams as nodes and their interactions (passes, duels, or competitive comparisons) as edges, it becomes possible to quantify and visualize the structural organization of the game. This approach has enabled, for instance, the measurement of player centrality in passing networks, the identification of team-specific tactical patterns, and the assessment of a playing system’s robustness to disruptions such as injuries, providing insights that are not accessible through traditional analytical methods \cite{caicedo2020passing,ichinose2021robustness,gonccalves2017exploring,chacoma2022complexity,chacoma2025identification,li2025motif,yung2025using}.

In this work, we propose a specific application of network theory to address a central problem in sports analytics: the objective and multidimensional comparison of team performance. Our approach consists of defining a rating system that is not based solely on final match outcomes, but on a set of performance metrics derived from event data.
This set of metrics are defined and justified in the following section.
To this end, we construct one network per metric, where nodes represent teams and weighted edges encode their relative performance in direct matchups. Applying the algebraic graph decomposition method \textit{HodgeRank} \cite{jiang2011statistical,perotti2025analysis} to these networks, allows us to infer a scalar rating for each team in each dimension of the game, thereby offering a nuanced, process-based comparison. 

The construction of these multidimensional, process-based ratings has direct practical implications. A single ranking, such as the league table, summarizes outcomes but can obscure the mechanisms underlying them. By decomposing performance into interpretable components (e.g., pressure, chance creation, possession times), our framework provides a structural diagnosis. For coaching staffs, this makes it possible to identify teams that accumulate points despite poor performance in key metrics—an indication of potential unsustainability—or, conversely, teams whose strong process-level performance does not translate into results, suggesting inefficiency in decisive moments. Such information is crucial for informed decision-making in areas such as player recruitment, tactical design, and opponent-specific preparation, allowing resources to be directed toward the aspects of play that truly determine long-term competitive advantage.
It is important to clarify the epistemic boundaries and the interpretative scope of the proposed framework. Our approach is strictly positioned within the realm of descriptive and diagnostic network analysis, rather than predictive or causal modeling. The correlations and alignments identified between our performance ratings and the official standings should not be interpreted as direct causal drivers of success, as tactical variables and match outcomes are intrinsically intertwined in a non-linear fashion. Instead, the true analytical value of this methodology lies in its ability to map macro-level structural patterns and, crucially, to evaluate the statistical residuals between process and outcome. Rather than viewing deviations from the official table as modeling errors, our framework treats these discrepancies as genuine physical indicators of efficiency, tactical bottlenecks, or stochastic factors (such as luck) that traditional score-based metrics naturally mask.

This article is organized as follows, to present the methodological framework and findings in a coherent manner. Section \ref{se:dym},  Data and Metrics, describes in detail the data collection process and sources, defines the eight performance metrics derived from match events, and explains the construction of the complex networks that represent competitive interactions between teams for each metric. Section \ref{se:theory}, Theoretical Framework, presents the algebraic foundations of the HodgeRank method, detailing how this graph decomposition technique is applied to infer ratings from the pairwise comparisons encoded in our networks. The main findings are presented in Section \ref{se:res}, Results, structured into five analytical subsections: (\ref{sse:rat}) an initial statistical analysis of the true rating and the validation of the hierarchical skill model; (\ref{sse:corr}) a global correlation analysis (Pearson coefficient) between the metric-based ratings and the true rating; (\ref{sse:ken}) a comparison of the resulting ordinal rankings using Kendall’s rank correlation coefficient; (\ref{se:incons}) an analysis of how structural inconsistencies affect the macroscopic ordering; and (\ref{sse:comp}) the proposal and optimization of a composite rating, obtained as a weighted linear combination of the most relevant metric ratings. Finally, the article concludes with Section \ref{se:concl}, Discussion and Conclusions, which integrates and interprets the overall results, discusses their implications, outlines the limitations of the study, and proposes concrete directions for future research.

\section{Data and metrics}
\label{se:dym}

\subsection{Metric collection}
\label{se:metricas}

In this work, we use the event data set provided by L. Pappalardo et al. in \cite{pappalardo2019public}.  
In that article, the authors manually annotate all matches from the 2017--2018 season of the major European football leagues: La Liga (Spain), Premier League (England), Serie A (Italy), Bundesliga (Germany), and Ligue 1 (France). For each match, they detect, classify, and localize in time and space all relevant events, including goals, shots, passes, corner kicks, fouls, among others.  
In the reference system used to locate events in time and space, $t$ denotes the elapsed time since the start of the match, the coordinate $x$ represents the distance relative to the goal defended by the team generating the event, and the coordinate $y$ represents the distance relative to the right touchline. Spatial coordinates are expressed as percentages of the pitch length, such that, for example, $x=0$, $x=50$, and $x=100$ correspond to the team’s own goal line, the midfield line, and the opponent’s goal line, respectively.

Within this framework, we define a ball possession interval (BPI) as the set given by a sequence of consecutive events generated by a single team. Note that each BPI contains information from only one team.  
We collect all BPIs from all teams in each league, and from these data we extract metrics that allow us to detect some of the tactical resources employed by teams within that temporal window of the match.  
The metrics considered in our analysis are based on those proposed by J. Fernández-Navarro in \cite{fernandez2018influence}. The study of these metrics was also employed in a previous work to identify characteristic football playing styles \cite{chacoma2025data}.  
Below, we describe each of these metrics in detail:

\begin{enumerate}
    \item {\it Direct play.} Each time a pass or a free kick occurs within a BPI, we measure the average velocity in the attacking direction, defined as the ratio between the distance traveled by the ball along the $x$ axis and the elapsed time. For each BPI, we retain the maximum value.  
    This metric allows us to quantify how directly the team moves the ball toward the opponent’s goal.

    \item {\it Counterattack.} Given two consecutive events within a BPI, if the first event occurs at $x_1<40$ and the second at $x_2>60$ with a temporal difference $\Delta t$, the velocity is reported as $v=\frac{x_2-x_1}{\Delta t}$. Otherwise, a value of $0$ is reported.  
    This metric measures how quickly a team transitions from a defensive position in its own half to an offensive position in the opponent’s half.

    \item {\it Build up.} If within a BPI it is verified that $\bar{x}>60$, i.e., the possession develops predominantly in the opponent’s half, the total possession time is reported. Otherwise, a value of $0$ is reported. This metric captures possession time in situations where the team strongly occupies the opponent’s territory.  

    \item {\it Midfield play.} If within a BPI it is observed that $\bar{x}\geq40$ and $\bar{x}\leq60$, i.e., the possession develops predominantly in the central area of the pitch, the total possession time is reported; otherwise, a value of $0$ is reported. The purpose of this variable is to measure the time the team spends in the midfield zone.

    \item {\it Flow rate.} For each BPI satisfying $\bar{x}\geq50$, we compute the temporal differences between all consecutive events and calculate their mean value, $\bar{dt}$. The metric is then defined as $1/\bar{dt}$. In this way, this metric provides a measure of how quickly the team circulates the ball in the opponent’s half.

    \item {\it Crossing.} If a crossing event is observed within a BPI, a value of $1$ is reported; otherwise, a value of $0$ is reported. This metric is used to count attempts to reach the penalty area through aerial play.

    \item {\it Pressure point.} For each BPI, we take the first event and extract its $x$ coordinate, i.e., the position at which the team starts its possession. This allows us to assess whether the team is recovering the ball in its own half, the midfield zone, or the opponent’s half.
 
    \item {\it Shots.} If a ``Shot'' event is recorded within a BPI, a value of $1$ is reported; otherwise, a value of $0$ is reported. This metric allows us to count the number of shots on goal produced by each team.
    
\end{enumerate}

For our analysis, all BPIs with fewer than 3 events and with a total duration shorter than 2 seconds were discarded. The rationale behind this choice is to exclude brief, transient recoveries and retain only consolidated possessions.  
It is important to emphasize that the selection of these eight metrics does not imply a claim of an exhaustive or universally optimal set for football tactical analysis. Instead, they represent a robust and theoretically validated selection of possession-based indicators that can be rigorously extracted from the available event-level data. Fundamentally, our objective is to evaluate whether these specific quantifiers contain sufficient structural information to derive a coherent competitive ranking. By doing so, we aim to develop a performance proxy capable of capturing underlying tactical nuances—such as possession structure and territory dominance—that traditional score-based outcomes, like goals or points, naturally overlook. Furthermore, the inherent flexibility of the proposed framework ensures that as more granular datasets become available, additional tactical dimensions can be seamlessly integrated into the model without altering its core mathematical structure.

From the data collection process, a total of $215681$ BPIs were obtained.
Subsequently, we aggregated the information by match and by team, summing the values obtained for each metric. In this way, for example, the feature {\it Shots} quantifies the total number of shots on goal taken by a team in a given match.  
Similarly, the feature {\it Build up} quantifies the net amount of time during which a team sustained an attacking possession against its opponent in that match.
Finally, in a separate dataset we collected meta-data associated with each BPI, which are later used in the analysis: the team corresponding to each one, the league it belongs to, and the final position in the league table.

\subsection{Representation as weighted fully connected networks}

In what follows, we present our proposal to represent performance metrics in terms of networks.  
We define $M(i,j,g)$ as the performance metric corresponding to team $i$ when facing team $j$ in match $g$. For example, this quantity may represent the number of shots on goal taken by FC Barcelona, represented by $i$, when playing against Real Madrid, represented by $j$, in the first encounter $g$ of the Spanish league {\it La Liga}.  
In our dataset, all teams participated in a round-robin league format, facing each opponent twice: a first-leg match ($g_1$) and a second-leg match ($g_2$). Using the information from both encounters, we define an aggregated metric that summarizes the performance observed between the two teams over the course of the season:
\begin{equation} \label{eq:1}
M(i,j) = \sum_{g= g_1, g_2} M(i,j,g).
\end{equation}
In the previous example, $M(i,j)$ represents the total number of shots on goal taken by FC Barcelona against Real Madrid across both matches of the tournament.
By computing $M(i,j)$ for each pair of teams in a league $L$, it is possible to represent these performance relationships by means of a directed and weighted graph $G(L,M)$, whose weights are defined as,
\begin{equation} \label{eq:2}
f_{ij} = \frac{M(j,i) - M(i,j)}{\overline{M_L}} = M^\prime(j,i)-M^\prime(i,j).
\end{equation}
Here, $\overline{M_L}$ is the average calculated over the ensemble of metric values observed in league $L$.
Note that, in this representation, $f_{ij} < 0$ indicates that team $i$ outperformed team $j$ with respect to the metric under consideration, where the difference is scaled by the average value of the ensemble.

Within this framework, we construct a total of 40 ``performance'' graphs, corresponding to 8 graphs per league, each one associated with a different performance metric.
Note that since the studied leagues follow a round-robin tournament format, the constructed graphs are fully connected.

\section{Theoretical Framework}
\label{se:theory}

In the previous section, we defined the performance graphs. The nodes in these graphs represent the teams of a particular league, the links express comparative relationships with respect to a given metric, and the weights quantify this comparison scaled by the ensemble's average value. In this section, we demonstrate how to apply HodgeRank theory to extract a set of ratings from these graphs, thereby defining the hierarchy of each league within the context of each specific metric.

To infer ratings $w_i$ from given comparison values $f_{ij}$, we seek to minimize the sum
\begin{equation}
\chi(w)
=
\sum_{ij}
a_{ij}\,
|f_{ij}-(w_j-w_i)|^2,
\end{equation}
with respect to the vector $w$ of components $w_i$.
Here, $a$ denotes the adjacency matrix of the network of matches between teams.
In the present work the network is fully connected, so $a_{ij}=1$ for $i\neq j$ and $a_{ii}=0$.

The edge weights are explicitly defined via the antisymmetric formulation $f_{ij} = M'(j,i) - M'(i,j)$, where $M'(i,j)$ represents the performance metric of team $i$ relative to team $j$, scaled by the ensemble's average. 
From a discrete calculus perspective, this antisymmetric structure is a
strict theoretical requirement of the HodgeRank method. The linear difference chosen here represents the most parsimonious operator to map edge flows to node potential differences, ensuring that the subsequent least-squares optimization is mathematically consistent and well-defined. 
It is worth noting that alternative non-linear formulations—such as applying log-transformations to handle heavy tails or utilizing local relative differences like $(M(j,i) - M(i,j))/(M(j,i) + M(i,j))$—were computationally tested. These alternative definitions yielded qualitatively identical rankings, proving the robustness of the framework. The global ensemble-average scaling was ultimately selected as it systematically minimized the network's residual inconsistencies while preserving the straightforward physical interpretation of potential gradients.

From a physical perspective, the ratings $w_i$ can be interpreted as potentials defined on the nodes, while the given weights $f_{ij}$ play the role of forces acting along the links or edges.
Under this interpretation, minimizing $\chi$ corresponds to finding the potentials $w_i$ whose differences $w_j-w_i$ reproduce the forces $f_{ij}$ as closely as possible in the least-squares sense.
This formulation effectively captures the relative competitive dynamics inherent to a round-robin tournament while preserving a clear physical analogy of potential gradients.

To infer the ratings $w_i$ efficiently, Jiang et al.~\cite{jiang2011statistical} introduced HodgeRank using the theory of discrete calculus on Higher Order Networks.
This theory interprets $f$ as a 1-cochain and finds $w$ through the Hodge decomposition~\cite{grady2010discrete,lim2020hodge,bianconi2021higher},
\begin{equation}
f = g \oplus s \oplus h ,
\end{equation}
where $g$, $s$, and $h$ represent gradient, curl, and harmonic components respectively.
In this setup, $w$ is a 0-cochain and is related to the gradient by $g=d_0 w$, where the coboundary operator $d_0$ is the discrete gradient operator.
This decomposition can be efficiently found by solving a pair of singular linear systems~\cite{perotti2025analysis}.

\section{Results}
\label{se:res}

\subsection{Statistics of the true rating}
\label{sse:rat}

Given that this work introduces a performance-based rating, it is essential to first analyze the statistics of the traditional system—namely, the ranking derived from accumulated points.
Therefore, the purpose of this section is to define a statistical model for the probability distribution of the true rating, $R_T$, defined as the total number of points obtained by a team over the course of a league season.
$R_T$ is a stochastic variable that depends on the number of matches won, drawn, and lost by teams in the league.
In football leagues, a team is awarded 3 points for a win, 1 point for a draw, and 0 points for a loss.
Within this framework, $R_T$ can be expressed in terms of the stochastic variables $W$ and $D$, which represent the number of matches won and drawn, respectively,
\begin{equation}
R_T = 3W + D.
\end{equation}
Let $n$ denote the total number of matches in the season. We model $W$ using a binomial distribution,
\begin{equation}
W \sim Bin( n | p_w ),
\end{equation}
where $p_w$ is the probability of winning a match. Conditional on the number of wins $W=w$, the number of draws follows,
\begin{equation}
D \mid W=w \sim Bin\!\left( n-w \,\bigg|\, \frac{p_d}{1-p_w} \right),
\end{equation}
where $p_d$ is the probability of drawing a match.
Since teams exhibit different competitive levels, each team is characterized by intrinsic probabilities $p_w$ and $p_d$ of winning and drawing, respectively.
We model these probabilities using a hierarchical approach: each team is endowed with a pair of latent skills $\boldsymbol{\eta} = (\eta_w, \eta_d)^T$, corresponding to win and draw abilities, which follow a bivariate normal distribution,
\begin{equation}
\boldsymbol{\eta} \sim \mathcal{N}\left(\boldsymbol{\mu}, \boldsymbol{\Sigma} \right),
\end{equation}
where,
\begin{equation}
\boldsymbol{\mu} = ( \mu_w, \mu_d)^T,
\quad
\boldsymbol{\Sigma} =
\begin{pmatrix}
\sigma_w^2 & \rho \sigma_w \sigma_d \\
\rho \sigma_w \sigma_d & \sigma_d^2
\end{pmatrix},
\end{equation}
here, $\boldsymbol{\mu}$ is the center of the distribution and $\boldsymbol{\Sigma}$ is the covariance matrix. In this framework, $\rho$ represents the correlation between win and draw abilities.
The probabilities are obtained through a softmax transformation,
\begin{equation}
\begin{aligned}
p_w &= \frac{e^{\eta_w}}{1 + e^{\eta_w} + e^{\eta_d}},\\
p_d &= \frac{e^{\eta_d}}{1 + e^{\eta_w} + e^{\eta_d}}.
\end{aligned}
\end{equation}
This transformation ensures that $p_w + p_d \leq 1$ and that the model is identifiable by implicitly fixing $\eta_l = 0$ for losses.
With these elements, we can write the joint distribution of wins and draws by marginalizing over the random effects $\boldsymbol{\eta}$,
\begin{equation} \label{integ}
\begin{split}
P( W=w, D=d ) = \int_{\mathbb{R}^2} & Bin( n \mid p_w(\boldsymbol{\eta}) ) \\
& \times Bin\!\left( n-w \,\bigg|\, \frac{p_d(\boldsymbol{\eta})}{1-p_w(\boldsymbol{\eta})} \right) \\
& \times \phi( \boldsymbol{\eta} \mid \boldsymbol{\mu}, \boldsymbol{\Sigma} ) \, d \boldsymbol{\eta}.
\end{split}
\end{equation}
Note that the integral in Eq.~\ref{integ} does not admit a closed-form solution; therefore, we evaluate it using Monte Carlo methods.
With these ingredients, we can finally write the theoretical probability distribution of the true rating as
\begin{equation} \label{theo}
P(R_T = r_T) = \sum_X P( W=w, D=d ),
\end{equation}
where the sum runs over the set of configurations $X$ defined by the constraints:
\begin{equation}
X = \{ (w, d) \in \mathbb{N}_0^2 : 3w + d = r_T, w + d \le n \}.
\end{equation}

\begin{figure}[t!]
\centering
\includegraphics[width=1.\textwidth]{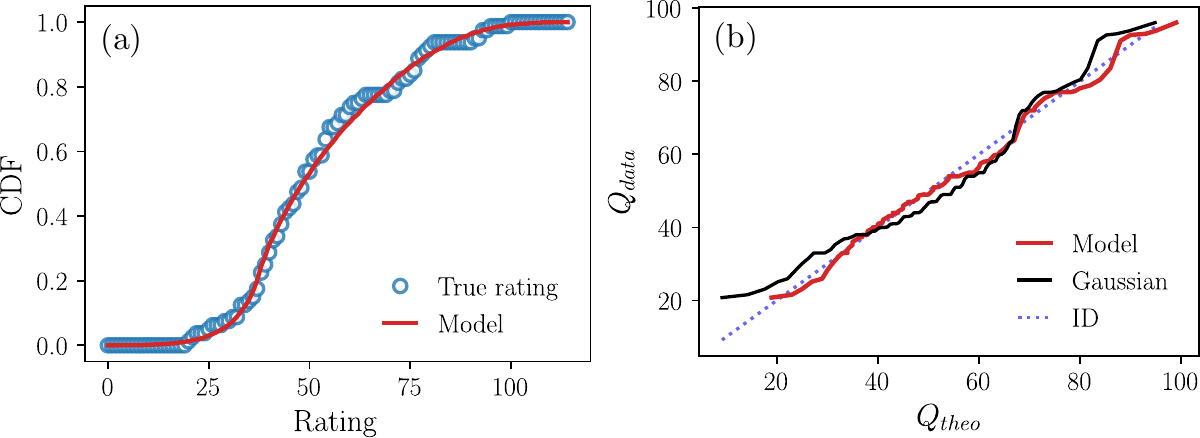}
\caption{Statistics of the true rating. 
(a) Comparison between the cumulative distribution function (CDF) of the true rating values and the distribution obtained from the proposed model. 
(b) Relationship between the empirical quantiles and those obtained from the model. For reference, a comparison with a Gaussian distribution with mean and standard deviation equal to those of the data is also shown.}
\label{fi:stats}
\end{figure}

The degrees of freedom provided by the parameters $\mu_w$, $\mu_d$, $\sigma_w$, $\sigma_d$, and $\rho$ allow the theoretical curve in Eq.~\ref{theo} to be fitted to the empirical distribution.
The fit was performed using the dataset associated with the English, French, Italian, and Spanish leagues. For this analysis, we chose not to include data from the German league, as its competition involves a smaller number of teams (18), which leads to a slightly different distribution of true rating values compared to the other leagues, where a total of 20 teams participate.
The fitting procedure was carried out using the Nelder--Mead algorithm, minimizing the RMSE between the empirical and theoretical CDFs.
Subsequently, using the optimal parameters, we performed parametric bootstrapping (1000 replicas) to estimate the mean values and uncertainties of the probabilities $p_w$ and $p_d$.
For the probability of winning, we obtained $\bar{p}_w = 0.366$ ($SD = 0.010$) with a $95\%$ confidence interval of $[0.3527, 0.3798]$.
Notice, the mean value of a given metric is denoted by $\bar{x}$, and $SD$ is the standard deviation.
Similarly, for the probability of drawing, we obtained $\bar{p}_d = 0.280$ ($SD = 0.020$) with a $95\%$ confidence interval of $[0.2548, 0.3074]$.
To contrast these results with empirical data, we computed for each team $i$ the empirical probabilities of winning and drawing, $q_w$ and $q_d$, using match outcome information.
Defininig $N_m^i$, $N_w^i$ and $N_d^i$ as the number of matches, number of matches winned and number of matches draw of team $i$, then,
\begin{equation}
\begin{aligned}
q_w^{(i)} &= \frac{N_w^{(i)}}{N_m^{(i)}}, \\
q_d^{(i)} &= \frac{N_d^{(i)}}{N_m^{(i)}}.
\end{aligned}
\end{equation}
By computing the mean value and standard deviation over the set of all teams, we obtained for the probability of winning $\bar{q}_w = 0.379$ ($SD = 0.168$) with a $95\%$ confidence interval of $[0.1572, 0.7638]$, and for the probability of drawing $\bar{q}_d = 0.243$ ($SD = 0.081$) with a $95\%$ confidence interval of $[0.1046, 0.3954]$.
From these results, we observe the following: (i) the mean values are similar ($0.366$ vs.\ $0.379$ for wins, $0.280$ vs.\ $0.243$ for draws), which validates the center of the distribution; (ii) the empirical standard deviations $(0.168, 0.081)$ are substantially larger than those implied by the model $(0.01, 0.02)$, indicating that the model underestimates team-level heterogeneity.
Regarding the correlation parameter, the parametric bootstrap yielded $\bar{\rho} = -0.130$ ($SD = 0.07$) with a $95\%$ confidence interval of $[-0.3040, -0.0371]$. This result indicates the presence of a weak negative structural correlation between teams’ win ability $\eta_w$ and draw ability $\eta_d$. This suggests that teams with a higher propensity to win tend, on average, to exhibit a slightly lower propensity to draw.
In Fig.~\ref{fi:stats} (a), we show the CDF of the true rating together with the fitted model. The proposed model accurately captures the behavior of the empirical curve across the entire support. In Fig.~\ref{fi:stats} (b), we present a quantile--quantile plot comparing theoretical and empirical quantiles. For reference, we also include the relationship between the data quantiles and those associated with a Gaussian distribution with mean and standard deviation equal to the sample values, $\bar{R}_T = 52.45$ ($SD = 18.56$).
First, we observe that the proposed model reproduces well the behavior in the central-left region and provides an acceptable description of the right tail. The comparison with the Gaussian model reveals that the empirical distribution exhibits lighter left tails and heavier right tails relative to a Gaussian distribution, thereby uncovering the presence of asymmetry. The downward concavity observed in the central region of the plot further indicates that the median of the data is shifted to the left.

\subsection{Correlation between true and metric-based ratings}
\label{sse:corr}

In this section, we characterize the differences between the true rating, as given by the league point system, and the rating obtained from the HodgeRank method, which we refer to as the metric rating.
Since eight performance metrics are considered (see Section~\ref{se:metricas}), for each team we compute eight different metric rating values, which are then compared with the true rating.
It is important to emphasize that the true rating and the metric rating are expressed on different scales. The former is measured in terms of total points obtained, whereas the latter is expressed as a standardized score.
Nevertheless, this does not prevent a meaningful comparison, as the relevant information conveyed by these quantities is not their absolute value, but rather the relative differences between teams.
For this reason, in order to compare the behavior of the true rating and the metric rating, we standardize both variables and work with their corresponding {\it z-scores}.

\begin{figure}[t!]
\centering
\includegraphics[width=1.\textwidth]{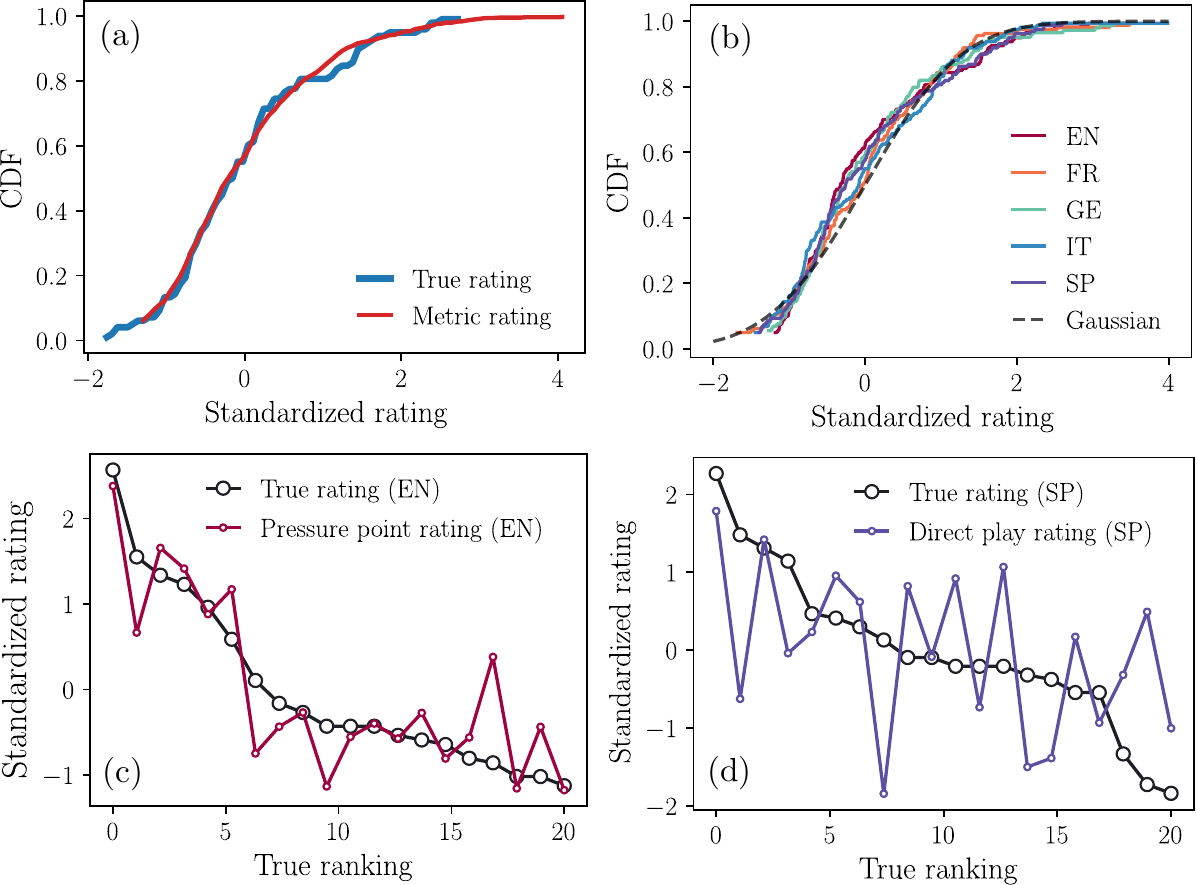}
\caption{Comparison between the true and metric-based rating. Since these quantities are expressed in different units, standardized values are used for comparison. 
(a) Comparison between the cumulative distribution functions associated with the true rating and the metric rating.
(b) Cumulative distribution function of the metric rating by league. The black dashed line represents, for reference, the CDF of a standard normal distribution.
(c) Comparison for the \textit{Pressure point} metric in the English league.
(d) Comparison for the \textit{Direct play} metric in the Spanish league.}
\label{fi:rating}
\end{figure}

In Fig.~\ref{fi:rating} (a), we show the CDF of the true rating computed over all teams and all leagues, together with the CDF of the metric rating computed over all teams, all leagues, and all eight metrics considered. For both curves, outlier values—defined as those above the quantile $Q_{99.7}$—were removed.
We observe a remarkable agreement across most of the range, with some differences appearing in the tails: the metric rating exhibits more extreme positive values relative to the mean, whereas the true rating shows slightly more extreme negative values.
In Fig.~\ref{fi:rating} (b), we show the distribution of metric rating values for each league. For this purpose, within each league we aggregate the metric rating values associated with all metrics and standardize them using the z-score.
As a reference, the CDF of a standard normal distribution is also shown as a black dashed line.
It can be observed that the curves associated with different leagues differ slightly, indicating league-dependent behavior. Moreover, in all cases we observe an accumulation of probability mass in the central region. This is compensated by lighter left tails and heavier right tails relative to the standard normal distribution.
On the other hand, in Fig.~\ref{fi:rating} (c) we plot the true rating and the metric rating as a function of the true ranking, for the case of the {\it Pressure point} metric in the English league. The purpose of this representation is to highlight team-by-team differences.
First, we note that the true rating decreases as the ranking increases. This decrease is not strictly monotonic, as plateaus may occur, but it is non-increasing by construction: the first-ranked team has a true rating greater than or equal to that of the second-ranked team, the second greater than or equal to the third, and so on until the last position.
For the metric rating curve, we also observe an overall decreasing trend, although not strictly. In qualitative terms, the metric rating follows the general trend of the true rating, while exhibiting small local deviations.
In Fig.~\ref{fi:rating} (d), we show an analogous comparison for the {\it Direct play} metric in the Spanish league. In this case, the metric rating curve does not display a clear decreasing trend and shows a strong decorrelation with respect to the true rating values.
These results indicate that, for a given league, the true rating may differ substantially or only marginally from the metric rating, depending on the specific metric used to infer the rating.

\begin{figure}[t!]
\centering
\includegraphics[width=1.\textwidth]{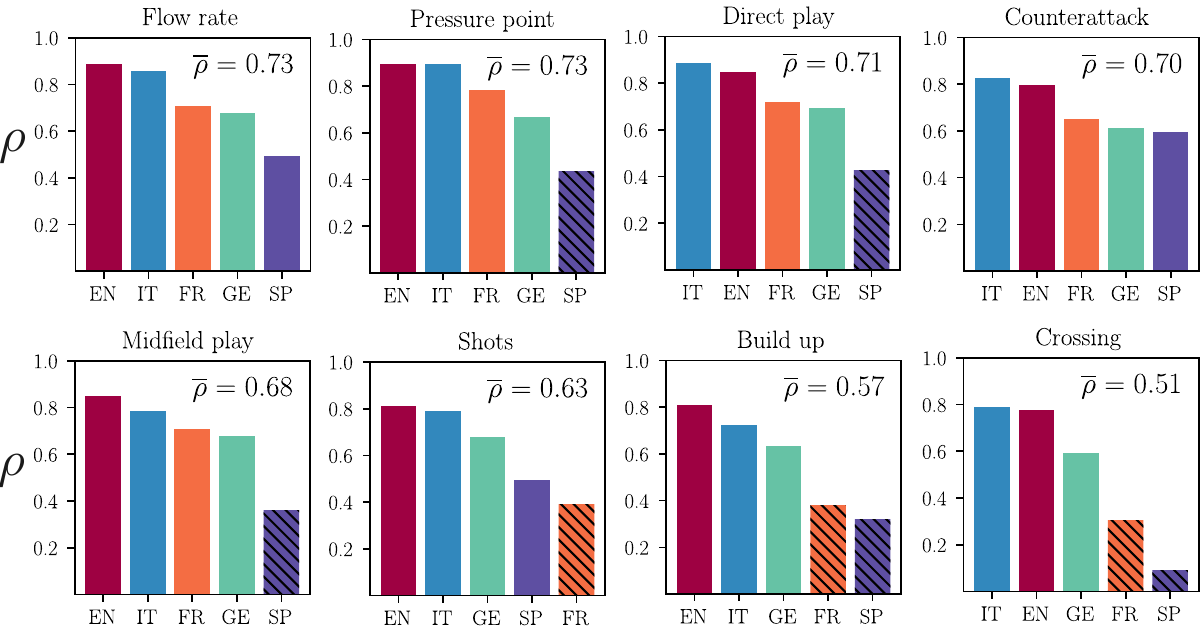}
\caption{Pearson correlation coefficient, $\rho$, between the true rating and the metric-based rating. 
Each panel corresponds to a performance metric, and the bars show the value of $\rho$ obtained for each league. 
Within each panel, bars are ordered in increasing order of $\rho$. Panels are also ordered in increasing order according to the average correlation value for each metric, $\overline{\rho}$. 
Bars with hatching indicate cases for which the $p$-value exceeds $0.05$. In these cases, no statistically significant correlation was found between the true rating and the metric rating.}
\label{fi:correlacion}
\end{figure}

Our next objective is to quantify the correlation between the true rating curve of a given league and the metric rating curves associated with each performance metric. Specifically, we aim to identify which metric yields a metric rating that best follows the trend of the true rating, and whether this relationship depends on the league.
To this end, we compute the Pearson correlation coefficient between the true rating of each league and the metric rating obtained from each metric.
The results are shown in Fig.~\ref{fi:correlacion}. Each panel corresponds to a different metric, and within each panel each bar represents a league. The panels are ordered from highest to lowest average correlation, and the bars within each panel are ordered from highest to lowest correlation value. Hatched bars indicate cases for which statistical significance cannot be established.
First, we observe that the Italian and English leagues exhibit the highest correlations across all metrics.
Second, we find that the French and German leagues alternate between third and fourth place among cases with intermediate correlation values for almost all metrics.
Finally, we observe that the Spanish league presents the weakest results, showing the lowest correlation values in 7 out of the 8 metrics considered. Nevertheless, it still exhibits strong correlations ($\rho > 0.4$) for the {\it Flow rate}, {\it Shots} and {\it Counterattack} metrics.

\subsection{Correlation between true and metric-based rankings}
\label{sse:ken}

\begin{figure}[t!]
\centering
\includegraphics[width=1.\textwidth]{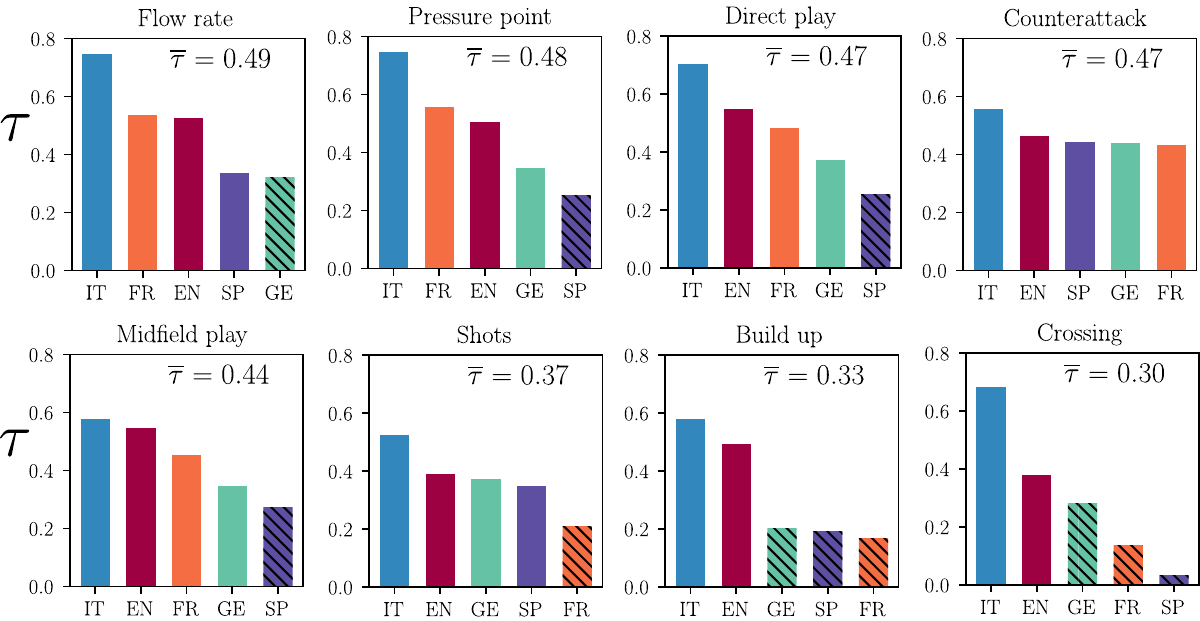}
\caption{Kendall rank correlation coefficient, $\tau$, between the true rankings and the rankings derived from performance metrics. 
Each panel corresponds to a performance metric, and the bars show the value of $\tau$ obtained for each league. 
Within each panel, bars are ordered in decreasing order of $\tau$. Panels are also ordered in decreasing order according to the average Kendall coefficient for each metric, $\overline{\tau}$. 
Bars with hatching indicate cases for which the $p$-value exceeds $0.05$. In these cases, the results do not provide sufficient evidence to assert the existence of a statistically significant association between the true ranking and the inferred ranking.}
\label{fi:rankings}
\end{figure}

In this section, we complement the analysis presented in the previous section by studying the correlation between rankings.
We define the \textit{true ranking} of a team as its final position in the league standings.
This ranking is obtained by ordering the teams in descending order according to the total number of points accumulated; in the event of a tie, the final ranking is determined by the goal difference (goals scored minus goals conceded).
Similarly, we define the \textit{metric ranking} by ordering the teams in a league in descending order according to the value of the \textit{metric rating}.
In this case, no ties were observed during the analysis.
To quantify the correlation between the two rankings, we employ Kendall’s rank correlation coefficient \cite{kendall1938new}, which is based on a nonparametric hypothesis test whose statistic, $\tau \in [-1,1]$, takes values close to one when the observations exhibit a similar ordering, and values close to zero when the orderings differ substantially.
Fig.~\ref{fi:rankings} presents the corresponding results.
Analogously to Fig.~\ref{fi:correlacion}, each panel corresponds to a metric, while each bar within a panel represents a league.
The panels are ordered from highest to lowest according to the average Kendall coefficient, and the bars within each panel are ordered in decreasing order according to their individual values.
Bars with hatching indicate cases for which statistical significance cannot be ensured.
First, we observe that the Italian league consistently leads all panels, exhibiting the highest values of Kendall’s coefficient.
The English league generally ranks second, although it is surpassed by the French league in two of the metrics.
By contrast, the German and Spanish leagues tend to occupy the lowest positions.
This behavior is, in principle, consistent with the results previously observed for the \textit{ratings}.
Second, we find that Kendall’s coefficients are, in general, slightly lower than the corresponding Pearson correlation coefficients.
This difference arises from the nonparametric nature of Kendall’s coefficient.
Importantly, this discrepancy does not imply a weak correlation, but rather reflects the increased sensitivity of rank-based measures to small fluctuations in regions of the table where point differences between teams are minimal.
While Pearson’s coefficient validates the global hierarchy of the league, Kendall’s coefficient highlights that the exact ordering of teams can be affected by small variations (noise), particularly in regions of the standings with a high density of points, that is, where multiple teams are tied or nearly tied.

By last, we compare the rankings obtained from this technique with those generated by the Bradley-Terry ranking model \cite{bradley1952rank}, which represents a standard approach in sports classification. In this work, we implement the variant with weight assignment based on the margin of victory, thereby ensuring that both methods operate with the same amount of input information as the HodgeRank.
Within this framework, we propose to treat the performance metrics as score lines. For any pair of competing teams, $i$ and $j$, their performances in a given metric are evaluated as $M^\prime(i,j)$ and $M^\prime(j,i)$ (see Eq.~\ref{eq:2}). The team that generates the larger value of $M$ wins the confrontation in that specific metric and is assigned a weight equal to the margin of victory, $|M^\prime(i,j)-M^\prime(j,i)|$, while the losing team receives a value of zero. Under this setup, the ranking is computed using the Bradley-Terry model for all metrics, and the Kendall-Tau correlation coefficient with respect to the true ranking ($\tau_{BT}$) is obtained. These values are compared against those derived from the Hodge Rank method ($\tau_{HR}$) in Table~\ref{tab:comp}.

At first glance, the results show that $\tau_{BT}$ exhibits values quite similar to $\tau_{HR}$ across all analyzed cases. To quantify these differences, we define $\Delta \tau = \tau_{HR}-\tau_{BT}$ and calculate its mean value and standard error, obtaining $\overline{\Delta \tau}= 0.005$ and $SE=0.017$. As can be observed, the confidence interval contains zero, therefore, we cannot conclude that, in global terms, either method provides a ranking that correlates significantly better or worse with the true ranking.
Nevertheless, by examining the extreme values of $\Delta \tau$, several interesting cases emerge for analysis. The instances where the difference exceeds the 90th percentile correspond to the metrics {\it Flow rate}, {\it Midfield play} and {\it Build up} all in the English league ($\tau_{HR} > \tau_{BT}$). 
Conversely, below the 10th percentile, we find the {\it Counterattack} in the Spanish league, {\it Shots} in the English league, and {\it Build up} in the Italian league ($\tau_{BT} > \tau_{HR}$).

\subsection{Analysis of network inconsistencies via the curl component}
\label{se:incons}

Thus far, we have analyzed how the inferred rating and ranking correlate with those obtained from the HodgeRank method. We observed that, depending on the analyzed league, the method captures the tournament structure better when using certain tactical performance metrics over others. Furthermore, we compared the HodgeRank method with the standard Bradley-Terry model and, using equivalent information, no statistically significant differences were detected in their macro-ordering capacity.
However, the HodgeRank method possesses a fundamental advantage over other traditional approaches: it allows for the isolation and quantification of local inconsistencies in dominance relationships, such as intransitive cyclic relationships (for instance, the scenario where team $A$ dominates $B$, $B$ dominates $C$, and $C$ dominates $A$). 
In the Hodge decomposition (see Eq. 4), the information associated with this cyclic behavior is encoded within the solenoidal (rotor) and harmonic components of each edge. Given that our networks are fully defined based on a round-robin competition system, mathematical theory dictates that the harmonic component vanishes due to the topology of the underlying simplicial complex. In this manner, the total net flow $f_{ij}$ on each edge is determined solely by the sum of two non-zero orthogonal components:
\begin{equation}
f_{ij} = g_{ij} + s_{ij}
\end{equation}
where $g_{ij}$ represents the gradient component, used to deduce the global rating $w_i$, and $s_{ij}$ corresponds to the solenoidal component, which absorbs the local circular inconsistencies associated with the network's triangles.
To quantify the global inconsistency level of the network under a specific metric $M$, we define the indicator $I$ as the ratio between the energy of the solenoidal component and the energy of the system's total flow:
\begin{equation}
I = \frac{\sum_{(i,j)} |s_{ij}|^2 }{\sum_{(i,j)} |f_{ij}|^2}.
\end{equation}

Under this formulation, the metric $I$ is strictly bounded between 0 and 1 due to the global orthogonality of the decomposition. A value of $I$ close to zero indicates that the network flow is purely hierarchical and that the analyzed metric exhibits low inconsistency relative to the global order induced by the ranking.

\begin{figure}[t!]
\centering
\includegraphics[width=1.\textwidth]{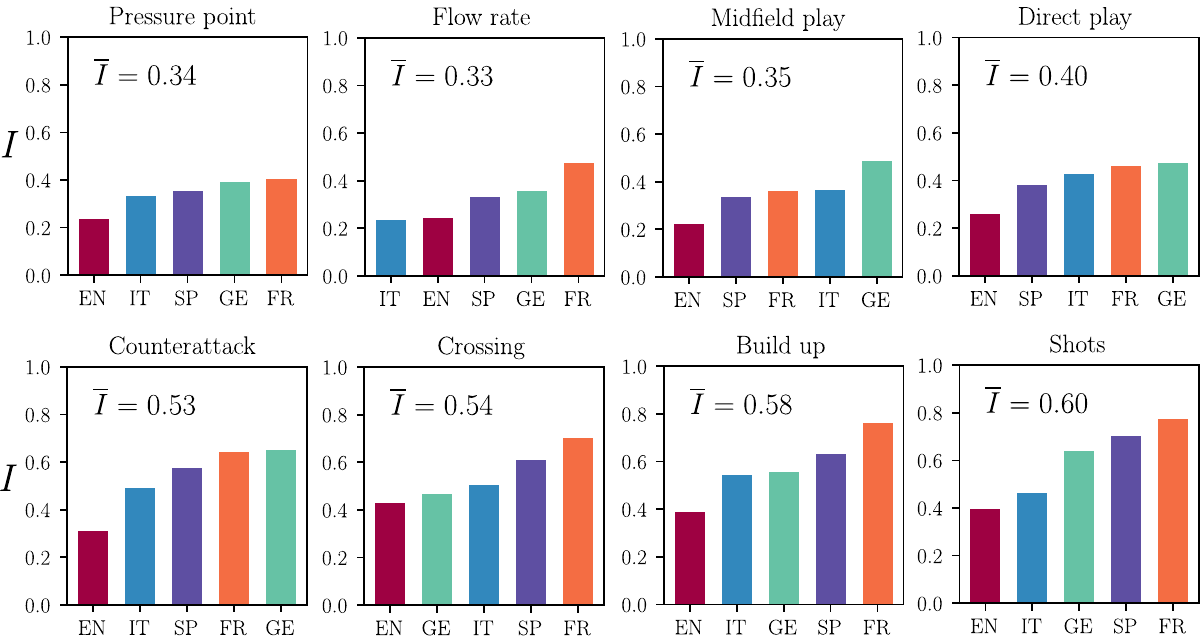}
\caption{
Inconsistency indicator, $I$. Each panel corresponds to a specific performance metric, with bars representing the $I$ values obtained for each league. Within each panel, the bars are arranged in ascending order with respect to $I$. The panels are also ordered ascendingly based on the average value per metric, $\overline{I}$.
}
\label{fi:I1}
\end{figure}

For each studied metric and league, the global inconsistency indicator $I$ was computed. The results are presented in Fig.~\ref{fi:I1}, where each panel groups the values corresponding to a specific tactical metric, and the bars represent the observed value of $I$ for each league. In order to facilitate the visualization of hierarchical patterns, the bars within each panel are sorted in increasing order (from lowest to highest inconsistency), while the panels themselves are sorted according to the average value of $I$ for each metric.
When analyzing the global behavior of the leagues, it is observed that the English league presents the lowest inconsistency across all analyzed variables, reaching its minimum value in the {\it Midfield play} metric. In contrast, the French league exhibits the highest levels of circularity, positioning itself as the most inconsistent league in 5 out of the 8 studied metrics. For its part, the German league shows a marked inconsistency in the {\it Counterattack} and {\it Shots} variables, where $I > 0.6$. Nonetheless, it records relatively low values in {\it Flow rate} and {\it Pressure point}. The Italian league, on the other hand, exhibits an intermediate and regular behavior, with its best performance in {\it Flow rate} and the greatest distortion in the {\it Build up} metric ($I > 0.5$). Finally, the Spanish league shows considerable dispersion, exceeding the threshold of $I > 0.5$ in half of the analyzed cases and reporting its highest level of consistency in {\it Midfield play}.

\begin{figure}[t!]
\centering
\includegraphics[width=1.\textwidth]{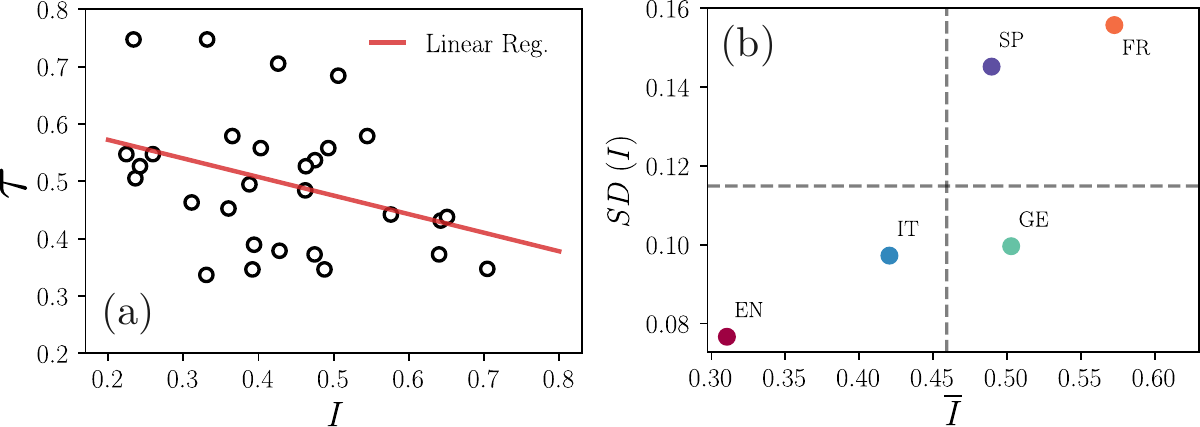}
\caption{
(a) Relationship between the inconsistency coefficient, $I$, and the Kendall rank correlation coefficient, $\tau$, across all studied metrics and leagues. The linear regression highlights how inconsistency impacts the correlation between the network-derived ranking and the true rating.
(b) Relationship between the mean inconsistency per league, $\overline{I}$, and its dispersion, quantified via the standard deviation, $SD(I)$. The dashed lines indicate the center of the data distribution for reference.
}
\label{fi:I2}
\end{figure}

In Fig.~\ref{fi:I2}(a), the global inconsistency indicator $I$ is contrasted against the Kendall $\tau$ coefficient for each metric and league, discarding those cases that are statistically non-significant ($p\text{-value} > 0.05$). A least-squares linear fit on this scatter plot yielded a slope of $\beta = -0.32$, quantitatively confirming the observed decreasing trend. 
Likewise, the Pearson correlation coefficient resulted in $\rho = -0.36$, evidencing the presence of a moderate negative correlation. This behavior empirically confirms how the emergence of local cyclic dynamics systematically degrades the capacity of the topological gradient to reconstruct the macro order of the competition. Those tactical performance variables that present lower energy dissipation in the solenoidal component are the ones that effectively filter out the volatility of the game, resulting in classifications that are better aligned with the real ranking of the standings table.

Lastly, Fig.~\ref{fi:I2}(b) examines the relationship between the average inconsistency of each league and its corresponding standard deviation. As a reference, the dashed lines indicate the center of the data distribution, allowing us to categorize the structure of the competitions into four conceptual quadrants, of which only three are occupied in our sample. 
In the low-inconsistency and low-dispersion quadrant, we find the English and Italian leagues. These competitions can be cataloged as highly hierarchical. The behavior of their teams is generally predictable, dominance relationships tend to respect the global ordering, and a low incidence of local tactical surprises is recorded.
In the zone of high inconsistency and low dispersion, we find the German league. In this case, the tournament can be thought of as behaving like a generalized cyclic system where teams tend to defeat each other in closed loops. Here, inconsistency is not an isolated event, but rather appears to be an intrinsic and uniform property of the league's ecosystem.
Finally, in the high-inconsistency and high-dispersion quadrant, we find the French and Spanish leagues, which can be associated with a regime of localized chaos. In these tournaments, teams with extreme volatility seem to coexist with others of high predictability, which shifts the statistical estimators toward boundary values and widens the variance of the system.

To conclude, it is important to remark that, beyond its impact on ranking precision, the quantification of global and local inconsistencies also offers an interesting diagnostic tool for sports analysis. 
In particular, this approach highlights the ``competition style'' of each league, allowing for the objective identification of whether a championship is dominated by a predictable hierarchy, a homogeneous parity where anyone can beat anyone, or isolated pockets of tactical volatility. This opens new avenues for evaluating collective behavior in professional football based on its intrinsic statistical properties.

\subsection{Composite rating based on performance metrics}
\label{sse:comp}

In this subsection, we develop a Composite Rating (CR) for each league, defined as a linear combination of the metric ratings obtained from performance metrics. The aim is to provide a more comprehensive and general assessment of team quality.
Let $n$ denote the number of teams in a league and $m$ the number of performance metrics. We define $\boldsymbol{X}^{n \times m}$ as the matrix whose rows contain the metric rating values obtained for each team. Similarly, we define $\boldsymbol{Y}^{n \times 1}$ as the vector containing the true ratings of each team in the league. Within this framework, our goal is to determine the weight vector $\boldsymbol{\psi}^{m \times 1} = (\psi_1, \ldots, \psi_m)^T$ and the intercept $\boldsymbol{b}$ such that
\begin{equation}
\boldsymbol{X} \boldsymbol{\psi} + \boldsymbol{b} = \boldsymbol{Y}.
\end{equation}
To solve this system by prioritizing both predictive power and parsimony, we chose to utilize a Lasso regression (Least Absolute Shrinkage and Selection Operator). This method incorporates a regularization term into the loss function that penalizes model complexity or, in other words, favors parsimony. The algorithm seeks to minimize the following cost function:
\begin{equation}\label{loss}
\min_{\omega} \left( \sum_{i=1}^{n} (y_i - \hat{y}_i)^2 + \alpha \sum_{j=1}^{p} |\psi_j| \right).
\end{equation}
The first term in Eq.~\ref{loss} corresponds to the Mean Squared Error (MSE), while the second term expresses the effect of $L_1$ regularization. In this framework, if the control parameter $\alpha$ is equal to zero, we recover an ordinary linear regression. On the other hand, if this parameter tends toward infinity, the algorithm will force all coefficients to zero. In this extreme scenario, the independent variables play no role, and the model is governed solely by the intercept. With intermediate values of $\alpha$, the algorithm nullifies the least explanatory variables, removing noise and favoring a simplified model that minimizes the generalization error.
To rigorously tune the value of $\alpha$, we employ the k-fold cross-validation technique with $k=5$. The algorithm evaluates a set of 100 candidate $\alpha$ values; for each, it randomly divides the dataset into 5 subsets (folds) of equal size. The model is trained using 4 of these subsets and validated with the remaining one. This procedure is repeated cyclically: in the first iteration, subset 1 acts as the validation set; in the second, subset 2 does so, and so on. For each of the 5 iterations, the MSE associated with the validation process is calculated. By averaging these values, we obtain the estimator $\overline{MSE}$, which characterizes the model's performance for a given value of $\alpha$.
This procedure is repeated for each of the 100 candidate values along the regularization path. The value that minimizes the estimation error, $\alpha_{min}$, is chosen to define the corresponding model for each of the studied leagues.

From the Lasso fit in each case, the regression coefficients $\psi_i$ that minimize the generalization error were obtained. It is important to note that since the variables were previously standardized, the obtained coefficients represent the relative ``importance'' of each tactical dimension in predicting the final league score.
Table~\ref{tab:CR} presents the regression results, the analysis of which reveals critical areas for optimizing sporting performance. For the English league, the most relevant metrics in order of relative importance are {\it Flow rate}, {\it Pressure point}, {\it Build up}, and, to a lesser but significant extent, {\it Crossing} and {\it Counterattack}. Together, these descriptors explain nearly $90\%$ of the data variance ($R^2 \approx 0.88$).
The positive coefficient for {\it Flow rate} ($20.24$) suggests that teams increasing the speed of ball circulation significantly improve their chances of success. Conversely, the negative coefficient for {\it Build up} ($-13.92$) indicates that teams reducing elaboration times in the final quarter of the pitch (see metric definition in Section~\ref{se:metricas}) achieve better results. An increase in this metric penalizes the composite rating, suggesting a structural inefficiency in playing styles that are excessively elaborate for this particular league. Furthermore, the positive value for {\it Pressure point} reinforces the importance of high pressing as a determining factor for performance.

The cases of the French and German leagues are of particular interest, as in both systems just a few metrics are sufficient to explain more than $50\%$ of the variance. In the French league, the {\it Pressure point} metric predominates over the others, underscoring that recovering the ball in the opponent's half is the primary predictor of success. For the German league, it is observed that shots, direct play, and density in the center of the pitch are the best proxies for understanding performance.
In the Italian league, the results suggest that success is associated with an increase in {\it Pressure point}, {\it Flow rate}, and {\it Direct play}, along with a decrease in {\it Midfield play} and {\it Build up}.
This describes a successful team profile characterized by verticality and rapid movement, relegating prolonged control in the middle and attacking zones. 
Finally, in the Spanish league, positive performance is linked to a high transition speed from the midfield upward and effectiveness in counterattacks, while excessively crossing appears correlated with low performance.

\begin{figure}[t!]
\centering
\includegraphics[width=1.\textwidth]{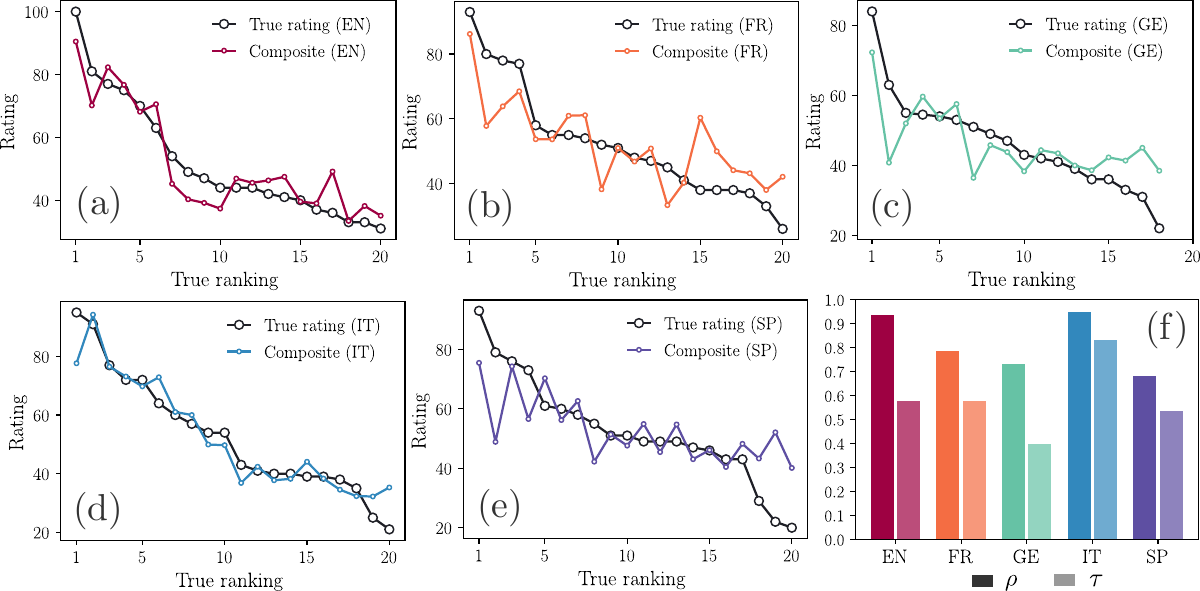}
\caption{Comparison between true and composite rating.
In panels (a), (b), (c), (d), and (e), we compare the true rating with the composite rating for each team in the English, French, German, Italian, and Spanish leagues, respectively. 
Panel (f) shows the values of the Pearson correlation coefficient, $\rho$, and the Kendall rank correlation coefficient, $\tau$, obtained for each league. 
Note that, in each case, the left bar corresponds to the value of $\rho$, while the right bar corresponds to the value of $\tau$.}
\label{fi:mv}
\end{figure}

The obtained composite rating can be compared with the true rating observed in the actual league table. In Fig.~\ref{fi:mv}, panels (a) through (e), we show this team-by-team comparison. In the case of the English league (Fig.~\ref{fi:mv} a), it is observed that for the top two teams, the true rating is higher than the composite rating. These teams exhibit statistical over-performance; that is, they obtained more points than their tactical indicators would predict, which could be attributed to external factors or favorable stochastic fluctuations (luck). Conversely, the team in third place shows under-performance: despite executing tactical metrics ``correctly'' according to the model, their real score is lower than expected, suggesting inefficiency in execution or a run of adverse results.
In general terms, the composite rating curves faithfully capture the trend of the real data, with the English and Italian leagues standing out in particular. To conclude, Fig.~\ref{fi:mv} (f) presents a comparison of the Pearson correlation coefficients ($\rho$) and Kendall's rank correlation ($\tau$). In all cases, the obtained values are higher than those achieved through individual metrics (compare with Fig.~\ref{fi:correlacion} and Fig.~\ref{fi:rankings}), validating the multivariate approach. We highlight the significant improvements in the Spanish and German leagues, where the aggregated model captures the system's complexity much more robustly than any isolated variable.

\section{Discussion and Conclusions}
\label{se:concl}

In this work, we present a unified methodological framework that integrates advanced football performance metrics, graph theory, and a robust algebraic method (HodgeRank) to generate team ratings and rankings.

As an initial step of the analysis, and with the aim of establishing a solid baseline for subsequent evaluation, we examined the statistical properties of the true ranking, defined by the total number of points obtained by teams in their respective leagues. To this end, we proposed a hierarchical model in which each team is characterized by three specific latent skill parameters: the abilities to win, draw, and lose. Analogously to rating systems such as the Elo method \cite{elo1978rating}, these skills define a probability distribution over match outcomes (win, draw, loss). Through a standard numerical fitting procedure, we estimated the corresponding theoretical probabilities, which showed satisfactory agreement with the observed empirical frequencies, thereby validating the adequacy of the proposed model.
For the generation of metric-based ratings, we collected eight performance metrics from an event-level database corresponding to the 2017--2018 season of the five major European leagues (England, France, Germany, Italy, and Spain). Based on these metrics, we constructed weighted undirected graphs, where nodes represent teams and edge weights encode relative performance information for a given metric. This procedure resulted in eight networks per league. The Hodge decomposition  was then applied to each graph, using the gradient component as the mechanism to derive a rating for each team. In this way, eight ratings per league (one per metric) were obtained and subsequently compared with the true ranking.
The relationship between these metric-based ratings and the true ranking was analyzed using Pearson’s correlation coefficient to assess global trends, complemented by Kendall’s rank correlation coefficient to evaluate the similarity in team ordering. The results revealed clear and consistent patterns, highlighting a strong dependence on both the league and the metric considered. In particular, the English and Italian leagues exhibited the highest levels of similarity between ratings derived from individual metrics and the true ranking, suggesting that, in these competitions, the selected metrics are individually effective predictors of overall performance. In contrast, the Spanish league showed the lowest levels of correspondence, with markedly weaker results across most metrics. This indicates that, for this league, the proposed metrics have limited predictive power and that alternative or complementary indicators may be required. These league-dependent differences further point to the existence of distinctive playing styles or competitive dynamics, a phenomenon already documented in the literature \cite{chacoma2025data}.
We also extended our topological analysis to evaluate the consistency of the inferred rankings. By quantifying the ratio between the solenoidal and total flow energies, we empirically confirmed that the emergence of local cyclic dynamics systematically limits the gradient component's capacity to reconstruct the macro-ordering of the competition, as evidenced by a moderate negative correlation with the true ranking. Crucially, our findings reveal that inconsistency is not a homogeneous artifact of the model, but rather a structural feature that characterizes the competition style of each league. The proposed framework successfully mapped these ecosystems into distinct regimes, distinguishing between highly hierarchical structures (such as the English and Italian leagues), homogeneous tactical parity defined by generalized loops (the German league), and pockets of localized chaos (the French and Spanish leagues). These results position the quantification of local and global inconsistencies as a robust diagnostic tool for sports analytics.
Finally, we defined for each league a composite rating, constructed as a weighted linear combination of the eight metric-based ratings. The weights were optimized through a fitting procedure designed to identify a parsimonious combination that best approximates the true ranking. This approach allowed us to obtain a more robust composite rating, significantly improving the predictive capability relative to individual metrics, with particularly notable improvements in the Spanish and German leagues. Moreover, this technique enabled us to quantify the relative importance of each metric within the context of each league, thereby providing an integrated view of team performance.
By optimizing independent LASSO regularization paths via cross-validation for each competition, we demonstrated that cross-league heterogeneity is an intrinsic statistical feature rather than a modeling artifact. Selecting the specific regularization parameters ($\alpha$) that minimize the cross-validation error ensures that the distinct subsets of selected metrics reflect robust, stable tactical fingerprints adapted to the specific competitive environment of each country. For instance, performance dimensions that act as structural pillars in one league are systematically discarded in another, formalizing the statistical quantification of different national playing cultures.
It is important to emphasize that this work does not propose a replacement for the official league table, whose competitive value and sporting legitimacy are not being questioned. We acknowledge that the inherently stochastic and unpredictable nature of football, which is often reflected in the official standings, constitutes an essential part of the sport’s appeal. Rather, the objective of our work is to provide a complementary analytical tool that, by uncovering the underlying statistical patterns of performance, offers a richer and more objective perspective on the competitive dynamics of a league. Such a tool could be used by coaching staffs to achieve a more comprehensive evaluation of teams’ true strength, beyond the contingencies of immediate match outcomes.
Regarding direct applications, the proposed methodological framework is generalizable to any head-to-head sport and to any set of performance metrics, thereby overcoming a key limitation of traditional rating models that rely exclusively on match outcomes (wins, draws, losses). By focusing on indicators of overall performance during the competitive process, our approach provides a potentially more stable evaluation with greater medium-term predictive power, as it is less sensitive to the randomness of isolated results.
Specifically, the proposed composite rating makes it possible to identify significant discrepancies between on-field performance and the points actually obtained. This capability is of substantial practical value: on the one hand, it can highlight teams with high ratings but low point totals, suggesting solid performance accompanied by bad luck or inefficiency in decisive moments. On the other hand, it can detect teams with low ratings that nonetheless occupy high positions in the standings, indicating performance sustained by favorable random factors. Identifying such statistically under- or over-performing teams provides coaches and analysts with valuable information to adjust strategies, manage expectations, and support technical decisions with a more solid quantitative basis.
Several avenues for future research emerge from this work. First, it would be essential to further investigate the inter-league differences observed. Incorporating contextual variables—such as financial resources, home advantage, or prevailing tactical models—would allow the development of a robust explanatory model describing how different football cultures are reflected in performance data and, consequently, in the derived ratings.
Such contextual analysis would not only help explain observed differences but also inform the development of more adaptive models. In this regard, a second priority direction would be to refine the composite rating through the incorporation of machine learning techniques and nonlinear models. These methods could dynamically optimize metric weights as a function of the specific context of each league or season, thereby overcoming the limitations of fixed-weight approaches.
Furthermore, given the model’s demonstrated ability to identify teams whose statistical performance diverges from their actual point totals, a natural and practically relevant extension would be the development of an early warning system. By integrating real-time data, such a system could continuously quantify a ``luck factor'' or efficiency in critical moments, providing coaching staffs with an analytical tool usable throughout the season.
Third, it would be of great interest to validate and extend the proposed framework to other team sports with similar confrontation dynamics. Generalization to disciplines such as basketball or hockey, using their specific performance metrics, would test the robustness of the approach and enable comparative inter-sport studies.
Finally, a significant methodological opportunity not yet explored in our analysis lies in fully exploiting the Hodge decomposition. Beyond the gradient and cyclic components used to generate the ratings, the harmonic component contain valuable structural information. 
This component, inherent to the topology of the graph, offers a complementary structural perspective. Although its contribution is null in a complete league graph, its analytical potential emerges when connectivity is redefined. For instance, if teams are connected not only through direct matchups but also through similarity in their metric profiles (forming a tactical similarity graph), the harmonic component could reveal natural groupings or communities of teams with similar playing styles. This would allow leagues to be characterized not merely as linear hierarchies, but as networks with possible tactical clusters, substantially enriching the description of competition beyond simple rankings.
The quantitative analysis and integration of these components would significantly enhance the characterization of a competition, adding a deeper interpretative layer to traditional classifications.

\section*{Acknowledgement} 
This work was partially supported by CONICET under Grant No. PIP 112 20200 101100 and SeCyT-UNC (Argentina).
This work used computational resources from UNC Supercómputo (CCAD), which is part of SNCAD, Argentina.


\begin{thebibliography}{10}

\bibitem{barthelemy2025fragility}
Marc Barthelemy.
\newblock Fragility of chess positions: Measure, universality, and tipping
  points.
\newblock {\em Physical Review E}, 111(1):014314, 2025.

\bibitem{Clauset2015safe}
A~Clauset, M~Kogan, and S~Redner.
\newblock Safe leads and lead changes in competitive team sports.
\newblock {\em Physical Review E}, 91(6):062815, 2015.

\bibitem{buldu2019defining}
Javier~M Buld{\'u}, Javier Busquets, Ignacio Echegoyen, and F~Seirul.~lo.
\newblock Defining a historic football team: Using network science to analyze
  guardiola’s fc barcelona.
\newblock {\em Scientific reports}, 9(1):13602, 2019.

\bibitem{Ribeiro2012anomalous}
Haroldo~V Ribeiro, Satyam Mukherjee, and Xiao Han~T Zeng.
\newblock Anomalous diffusion and long-range correlations in the score
  evolution of the game of cricket.
\newblock {\em Physical Review E}, 86(2):022102, 2012.

\bibitem{Chacoma2022simple}
Andr{\'e}s Chacoma and Orlando~V Billoni.
\newblock Simple mechanism rules the dynamics of volleyball.
\newblock {\em Journal of Physics: Complexity}, 3(3):035006, 2022.

\bibitem{Chacoma2023probabilistic}
Andr{\'e}s Chacoma and Orlando~V Billoni.
\newblock Probabilistic model for padel games dynamics.
\newblock {\em Chaos, Solitons \& Fractals}, 174:113784, 2023.

\bibitem{chacoma2025emergent}
Andr{\'e}s Chacoma and Orlando~V Billoni.
\newblock Emergent complexity in the decision-making process of chess players.
\newblock {\em Scientific Reports}, 15(1):23234, 2025.

\bibitem{zappala2022role}
Chiara Zappal{\'a}, Alessandro Pluchino, Andrea Rapisarda, Alessio~Emanuele
  Biondo, and Pawel Sobkowicz.
\newblock On the role of chance in fencing tournaments: An agent-based
  approach.
\newblock {\em Plos one}, 17(5):e0267541, 2022.

\bibitem{ibanez2018relative}
Sergio~J Ib{\'a}{\~n}ez, Aitor Mazo, Juarez Nascimento, and Javier
  Garc{\'\i}a-Rubio.
\newblock The relative age effect in under-18 basketball: Effects on
  performance according to playing position.
\newblock {\em PloS one}, 13(7):e0200408, 2018.

\bibitem{galeano2022using}
Javier Galeano, Miguel-{\'A}ngel G{\'o}mez, Fernando Rivas, and Javier~M
  Buld{\'u}.
\newblock Using markov chains to identify player’s performance in badminton.
\newblock {\em Chaos, Solitons \& Fractals}, 165:112828, 2022.

\bibitem{silva2016sports}
Pedro Silva, Lu{\'\i}s Vilar, Keith Davids, Duarte Ara{\'u}jo, and J{\'u}lio
  Garganta.
\newblock Sports teams as complex adaptive systems: manipulating player numbers
  shapes behaviours during football small-sided games.
\newblock {\em SpringerPlus}, 5(1):191, 2016.

\bibitem{ashford2021understanding}
Michael Ashford, Andrew Abraham, and Jamie Poolton.
\newblock Understanding a player’s decision-making process in team sports: a
  systematic review of empirical evidence.
\newblock {\em Sports}, 9(5):65, 2021.

\bibitem{martinez2020spatial}
Johann~H Mart{\'\i}nez, David Garrido, Jos{\'e}~L Herrera-Diestra, Javier
  Busquets, Ricardo Sevilla-Escoboza, and Javier~M Buld{\'u}.
\newblock Spatial and temporal entropies in the spanish football league: A
  network science perspective.
\newblock {\em Entropy}, 22(2):172, 2020.

\bibitem{cao2023football}
Xiaoxiang Cao, Xiaodong Zhao, Huan Tang, Nianchun Fan, and Fateh Zereg.
\newblock Football players’ strength training method using image processing
  based on machine learning.
\newblock {\em Plos one}, 18(6):e0287433, 2023.

\bibitem{yamamoto2024theory}
Ken Yamamoto, Seiya Uezu, Keiichiro Kagawa, Yoshihiro Yamazaki, and Takuma
  Narizuka.
\newblock Theory and data analysis of player and team ball possession time in
  football.
\newblock {\em Physical Review E}, 109(1):014305, 2024.

\bibitem{chacoma2020modeling}
A~Chacoma, Nahuel Almeira, Juan~Ignacio Perotti, and Orlando~Vito Billoni.
\newblock Modeling ball possession dynamics in the game of football.
\newblock {\em Physical Review E}, 102(4):042120, 2020.

\bibitem{chacoma2021stochastic}
A~Chacoma, N~Almeira, JI~Perotti, and OV~Billoni.
\newblock Stochastic model for football's collective dynamics.
\newblock {\em Physical Review E}, 104(2):024110, 2021.

\bibitem{chacoma2025data}
Andres Chacoma and Orlando~V Billoni.
\newblock Data-driven approach to defining football styles in major leagues.
\newblock {\em Chaos, Solitons \& Fractals}, 200:116926, 2025.

\bibitem{amichay2025characterizing}
Guy Amichay, Hugo Silva, Jo{\~a}o Brito, and Rui Marcelino.
\newblock Characterizing the spatial structures of competing football teams.
\newblock {\em Scientific Reports}, 15(1):35217, 2025.

\bibitem{caicedo2020passing}
Sergio Caicedo-Parada, Carlos Lago-Pe{\~n}as, and Enrique Ortega-Toro.
\newblock Passing networks and tactical action in football: A systematic
  review.
\newblock {\em International Journal of Environmental Research and Public
  Health}, 17(18):6649, 2020.

\bibitem{ichinose2021robustness}
Genki Ichinose, Tomohiro Tsuchiya, and Shunsuke Watanabe.
\newblock Robustness of football passing networks against continuous node and
  link removals.
\newblock {\em Chaos, Solitons \& Fractals}, 147:110973, 2021.

\bibitem{gonccalves2017exploring}
Bruno Gon{\c{c}}alves, Diogo Coutinho, Sara Santos, Carlos Lago-Penas, Sergio
  Jim{\'e}nez, and Jaime Sampaio.
\newblock Exploring team passing networks and player movement dynamics in youth
  association football.
\newblock {\em PloS one}, 12(1):e0171156, 2017.

\bibitem{chacoma2022complexity}
A~Chacoma, OV~Billoni, and MN~Kuperman.
\newblock Complexity emerges in measures of the marking dynamics in football
  games.
\newblock {\em Physical Review E}, 106(4):044308, 2022.

\bibitem{chacoma2025identification}
Andr{\'e}s Chacoma.
\newblock Identification and optimization of high-performance passing networks
  in football.
\newblock {\em Physical Review E}, 111(4):044313, 2025.

\bibitem{li2025motif}
Ming-Xia Li, Li-Gong Xu, and Wei-Xing Zhou.
\newblock Motif analysis and passing behavior in football passing networks.
\newblock {\em Chaos, Solitons \& Fractals}, 190:115750, 2025.

\bibitem{yung2025using}
Kate~KY Yung, Paul~PY Wu, Karen aus~der F{\"u}nten, Anne Hecksteden, and Tim
  Meyer.
\newblock Using a bayesian network to classify time to return to sport based on
  football injury epidemiological data.
\newblock {\em PloS one}, 20(3):e0314184, 2025.

\bibitem{bradley1952rank}
RA~Bradley and ME~Terry.
\newblock Rank analysis of incomplete block designs: {I}. {T}he method of paired comparisons.
\newblock {\em Biometrika}, 39(3/4):324--337, 1952.

\bibitem{jiang2011statistical}
Xiaoye Jiang, Lek-Heng Lim, Yuan Yao, and Yinyu Ye.
\newblock Statistical ranking and combinatorial hodge theory.
\newblock {\em Mathematical Programming}, 127(1):203--244, 2011.

\bibitem{perotti2025analysis}
Juan~I. Perotti.
\newblock Analysis of the inference of ratings and rankings in complex networks
  using discrete exterior calculus on higher-order networks.
\newblock {\em Phys. Rev. E}, 111:034306, Mar 2025.

\bibitem{pappalardo2019public}
Luca Pappalardo, Paolo Cintia, Alessio Rossi, Emanuele Massucco, Paolo
  Ferragina, Dino Pedreschi, and Fosca Giannotti.
\newblock A public data set of spatio-temporal match events in soccer
  competitions.
\newblock {\em Scientific data}, 6(1):236, 2019.

\bibitem{fernandez2018influence}
Javier Fernandez-Navarro, Luis Fradua, Asier Zubillaga, and Allistair~P.
  McRobert.
\newblock Influence of contextual variables on styles of play in soccer.
\newblock {\em International Journal of Performance Analysis in Sport},
  18(3):423--436, 2018.

\bibitem{grady2010discrete}
Leo~J. Grady and Jonathan~R. Polimeni.
\newblock {\em Discrete Calculus: Applied Analysis on Graphs for Computational
  Science}.
\newblock Springer, 2010.

\bibitem{lim2020hodge}
Lek-Heng Lim.
\newblock Hodge laplacians on graphs.
\newblock {\em SIAM Review}, 62(3):685--715, 2020.

\bibitem{bianconi2021higher}
Ginestra Bianconi.
\newblock {\em Higher-Order Networks}.
\newblock Elements in the Structure and Dynamics of Complex Networks. Cambridge
  University Press, 2021.

\bibitem{kendall1938new}
Maurice~G Kendall.
\newblock A new measure of rank correlation.
\newblock {\em Biometrika}, 30(1-2):81--93, 1938.

\bibitem{elo1978rating}
Arpad~E. Elo.
\newblock {\em The Rating of Chess Players, Past and Present}.
\newblock Arco Publishing, New York, 1978.

\end{thebibliography}

\pagebreak
\newpage

\begin{table}[htbp]
\centering
\caption{Kendall rank correlation values obtained by contrasting the true ranking with (i) the ranking derived from the HodgeRank method, $\tau_{HR}$, and (ii) the ranking derived from the Bradley-Terry method, $\tau_{BT}$. Results are shown for all metrics across all studied leagues. Values marked with an asterisk ($^*$) indicate that the correlation is not statistically significant ($p\text{-value} > 0.05$).}
\label{tab:comp}
\scriptsize 
\begin{tabular}{l *{10}{c}} 
\toprule 
\textbf{Metric} & \multicolumn{2}{c}{\textbf{England}} & \multicolumn{2}{c}{\textbf{France}} & \multicolumn{2}{c}{\textbf{Germany}} & \multicolumn{2}{c}{\textbf{Italy}} & \multicolumn{2}{c}{\textbf{Spain}} \\ 
\cmidrule(lr){2-3} \cmidrule(lr){4-5} \cmidrule(lr){6-7} \cmidrule(lr){8-9} \cmidrule(lr){10-11}
& $\tau_{HR}$ & $\tau_{BT}$ & $\tau_{HR}$ & $\tau_{BT}$ & $\tau_{HR}$ & $\tau_{BT}$ & $\tau_{HR}$ & $\tau_{BT}$ & $\tau_{HR}$ & $\tau_{BT}$ \\ 
\midrule 
Flow rate      & 0.53 & 0.43 & 0.54 & 0.51 & 0.32$^*$ & 0.35     & 0.75 & 0.84 & 0.34     & 0.25$^*$ \\ 
Pressure point & 0.51 & 0.42 & 0.56 & 0.49 & 0.35     & 0.29$^*$ & 0.75 & 0.72 & 0.25$^*$ & 0.29$^*$ \\ \addlinespace
Direct play    & 0.55 & 0.49 & 0.48 & 0.45 & 0.37     & 0.29$^*$ & 0.71 & 0.64 & 0.25$^*$ & 0.25$^*$ \\ 
Counterattack  & 0.46 & 0.48 & 0.43 & 0.35 & 0.44     & 0.19$^*$ & 0.56 & 0.57 & 0.44     & 0.55     \\ \addlinespace
Midfield play  & 0.55 & 0.44 & 0.45 & 0.49 & 0.35     & 0.24$^*$ & 0.58 & 0.58 & 0.27$^*$ & 0.31$^*$ \\ 
Shots          & 0.39 & 0.54 & 0.21$^*$ & 0.16$^*$ & 0.37 & 0.31$^*$ & 0.53 & 0.58 & 0.35     & 0.32$^*$ \\ \addlinespace
Build up       & 0.49 & 0.40 & 0.17$^*$ & 0.12$^*$ & 0.20$^*$ & 0.25$^*$ & 0.58 & 0.75 & 0.19$^*$ & 0.21$^*$ \\ 
Crossing       & 0.38 & 0.33 & 0.14$^*$ & 0.09$^*$ & 0.28$^*$ & 0.27$^*$ & 0.68 & 0.72 & 0.03$^*$ & -0.01$^*$ \\ 
\bottomrule 
\end{tabular}
\end{table}

\begin{table}[htbp]
\centering
\caption{No--null components of the composite rating by league based on Lasso regression.}
\label{tab:CR}
\begin{tabular}{lcc}
\toprule
\textbf{Metric} & \textbf{$\psi$} & \textbf{$R^2$} \\
\midrule
\multicolumn{3}{l}{\textbf{England}} \\ 
\quad Flow rate      & 20.24  & 0.88 \\ 
\quad Pressure point & 14.01  &       \\ 
\quad Build up       & -13.92 &       \\ 
\quad Crossing       & -2.52  &       \\ 
\quad Counterattack  & -1.75  &       \\ 
\addlinespace
\multicolumn{3}{l}{\textbf{France}} \\ 
\quad Pressure point & 11.35  & 0.62 \\ 
\quad Counterattack  & 1.13   &       \\ 
\addlinespace
\multicolumn{3}{l}{\textbf{Germany}} \\ 
\quad Shots          & 3.71   & 0.53 \\ 
\quad Direct play    & 3.12   &       \\ 
\quad Midfield play  & 2.68   &       \\ 
\quad Crossing       & 0.26   &       \\ 
\addlinespace
\multicolumn{3}{l}{\textbf{Italy}} \\ 
\quad Pressure point & 11.64  & 0.90 \\ 
\quad Flow rate      & 9.74   &       \\ 
\quad Midfield play  & -8.95  &       \\ 
\quad Direct play    & 8.32   &       \\ 
\quad Build up       & -6.32  &       \\ 
\quad Crossing       & 3.13   &       \\ 
\quad Shots          & 0.49   &       \\ 
\addlinespace
\multicolumn{3}{l}{\textbf{Spain}} \\ 
\quad Counterattack  & 6.19   & 0.46 \\ 
\quad Crossing       & -5.00  &       \\ 
\quad Shots          & 4.26   &       \\ 
\quad Flow rate      & 4.11   &       \\ 
\quad Midfield play  & -0.67  &       \\ 
\bottomrule
\end{tabular}
\end{table}


\end{document}